\documentclass{article}

 \usepackage[preprint]{neurips_2025}

\usepackage[utf8]{inputenc} 
\usepackage[T1]{fontenc}    
\usepackage{hyperref}       
\usepackage{url}            
\usepackage{booktabs}       
\usepackage{amsfonts}       
\usepackage{nicefrac}       
\usepackage{microtype}      
\usepackage{xcolor}         
\usepackage{graphicx}

\title{An in-silico integration of neurodevelopmental and dopaminergic views of schizophrenia}

\author{%
Xena Al-Hejji \\  
Department of Psychology\\
Mount Royal University\\
Calgary, Alberta, Canada \\
\And
Santina Duarte \\
Biology Department \\
Mount Royal University \\
Calgary, Alberta, Canada
\And
Jose Guillermo Gomez Castro \\
Department of Psychology\\
Mount Royal University\\
Calgary, Alberta, Canada \\
\And
Edgar Bermudez Contreras \\
OraQ AI \\
\texttt{edgar.bermudez@gmail.com } \\
\And
Eric Chalmers \\
Department of Math \& Computing \\
Mount Royal University \\
\texttt{echalmers@mtroyal.ca} \\
}

\begin{document}

\maketitle

\begin{abstract}
Deep reinforcement learning (DRL) algorithms have the potential to provide new insights into psychiatric disorders. Here we create a DRL model of schizophrenia: a complex psychotic disorder characterized by anhedonia, avoidance, temporal discounting, catatonia, and hallucinations. Schizophrenia’s causes are not well understood: dopaminergic theories emphasize dopamine system dysfunction, while neurodevelopmental theories emphasize abnormal connectivity, including excitation/inhibition (E/I) imbalance in the brain. In this study, we suppressed positive (excitatory) connections within an artificial neural network to simulate E/I imbalance. Interestingly, this is insufficient to create behavioral changes; the network simply compensates for the imbalance. But in doing so it becomes more sensitive to noise. Injecting noise into the network then creates a range of schizophrenic-like behaviours. These findings point to an interesting potential pathology of schizophrenia: E/I imbalance leads to a compensatory response by the network to increase the excitability of neurons, which increases susceptibility to noise. This suggests that the combination of E/I imbalance and neural noise may be key in the emergence of schizophrenic symptoms. We further notice altered response to reward prediction error in our model, and thus propose that E/I imbalance plus noise can account for both schizophrenia symptoms and dopamine system dysfunction: potentially unifying dopaminergic and neurodevelopmental theories of schizophrenia pathology.

\end{abstract}

\section{Introduction}

Schizophrenia is a debilitating psychotic disorder characterized by “positive” symptoms including hallucinations, and “negative” symptoms including disorganized or catatonic behavior, anhedonia, and blunted affect \cite{americanpsychiatricassociationDiagnosticStatisticalManual2022}. Up to 60\% of schizophrenic individuals experience negative symptoms, yet the pathogenesis of these symptoms is unclear and they often respond poorly to antipsychotics, a first-line treatment for the disorder \cite{correllNegativeSymptomsSchizophrenia2020}. Development of schizophrenia has been linked to genetic factors \cite{sullivanSchizophreniaComplexTrait2003} as well as environmental factors such as cannabis use during adolescence \cite{casadioCannabisUseYoung2011}, immigration \cite{cantor-graaeSchizophreniaMigrationMetaanalysis2005}, urban living \cite{pedersenAreCausesResponsible2006}, and prenatal exposure to infection \cite{brownPrenatalInfectionSchizophrenia2010}. However, despite decades of research the neural mechanisms of schizophrenia remain elusive, and there are several theories around schizophrenia’s underlying mechanisms.

\subsection{Theories of schizophrenia pathology}

Two prominent theories of schizophrenia pathology are the dopaminergic hypothesis and the neurodevelopmental hypothesis.

The dopaminergic hypothesis posits that positive symptoms are associated with excess dopamine subcortically, while negative and cognitive symptoms of schizophrenia are associated with deficient dopamine in the cortex \cite{abi-darghamWeStillBelieve2004}. For years, this remained the primary hypothesis largely due to the efficacy of dopamine receptor antagonistic antipsychotics as a treatment for schizophrenia. However, antipsychotics are only effective in treating positive symptoms of schizophrenia and have very little effect on negative or cognitive symptoms \cite{glausierDendriticSpinePathology2013}, \cite{lewisNeuroplasticityNeocorticalCircuits2008}. Therefore, although dopamine dysregulation does appear to play an important role in schizophrenia, it does not fully account for all symptoms; this has prompted search for alternative, more comprehensive hypotheses.

The neurodevelopmental hypothesis attributes schizophrenia to abnormal brain development through adolescence \cite{selemonSchizophreniaTaleTwo2015}, \cite{marencoNeurodevelopmentalHypothesisSchizophrenia2000}. Studies have found that schizophrenia is associated with gray matter loss, synapse loss, and dendritic spine loss \cite{penzesDendriticSpinePathology2011}: it is thought that excessive synaptic pruning during adolescence, or underproduction of dendritic spines in early childhood \cite{penzesDendriticSpinePathology2011, moyerDendriticSpineAlterations2015} somehow cause the onset of schizophrenia in adolescence or early adulthood \cite{patelSchizophreniaOverviewTreatment2014}. Regardless of how this pathology emerges, there is a consistent observation of impaired synaptic connectivity in schizophrenia.

More recent hypotheses have attempted to reconcile the neurodevelopmental hypothesis with the dopamine hypothesis. The integrated hypothesis of schizophrenia combines the ideas of abnormal brain development with disruptions in dopamine systems, proposing that excessive synaptic pruning disrupts the excitation-inhibition (E/I) balance, the equilibrium between excitatory and inhibitory inputs, and that this E/I disruption leads to elevated dopamine release \cite{chafeeUnmaskingSchizophreniaSynaptic2022, howesIntegratingNeurodevelopmentalDopamine2022}. There is a growing body of evidence supporting the presence of an E/I imbalance in schizophrenia, with studies finding altered excitatory and inhibitory activity at the molecular, cellular, and circuit level \cite{liuSelectiveReviewExcitatoryInhibitory2021, gaoChapter22Synaptic2016}. E/I balance plays a crucial role in efficient information processing at the level of neurons, synapse, circuits, and networks \cite{gaoChapter22Synaptic2016}, and it is thought that this E/I imbalance, triggered by aberrant synaptic pruning, underlies the symptoms associated with schizophrenia \cite{howesIntegratingNeurodevelopmentalDopamine2022}. Moreover, the cortical E/I imbalance may dysregulate neurons that project from the frontal cortex to key regions such as the striatum, elevating dopamine activity and ultimately resulting in psychotic symptoms \cite{howesIntegratingNeurodevelopmentalDopamine2022, howesSynapticHypothesisSchizophrenia2023}.

It is difficult to directly test the effects of an E/I imbalance \textit{in vivo}, since the origin of the E/I imbalance in schizophrenia is unclear, and it cannot be induced/reversed easily in animal models. Here we devise a way to test it \textit{in silico} using Reinforcement Learning techniques; collecting experimental data in a computational setting that is very difficult to collect in a biological one.

\subsection{Reinforcement Learning}

Because of the close analogy between DRL and biological reward-based learning, DRL is starting to be used in significant neuroscientific modeling and hypothesis creation, though much potential remains untapped \cite{botvinickDeepReinforcementLearning2020}. In particular, it is a promising technique for modelling schizophrenia, since it accounts for the effects of dopamine (emphasized by the dopaminergic hypothesis) and the role of neural connectivity (emphasized by the neurodevelopmental hypothesis). Such a model could potentially suggest how the two theories might be reconciled.

\subsection{Altered signal-to-noise ratio in schizophrenia}

Neural noise and signal-to-noise ratio (SNR) do not feature in the dopaminergic hypothesis’ or neurodevelopmental hypothesis’ narratives of schizophrenia. However, various studies have observed altered SNR in schizophrenia, and the work in this paper finds noise to be a key ingredient in schizophrenia-like behavior.

Signal-to-noise ratio refers to the ratio between meaningful, stimulus-driven signals and spontaneous, stimulus-independent fluctuations in brain activity \cite{wintererSchizophreniaReducedSignaltonoise2000, waschkeStatesTraitsNeural2017, abbasiIncreasedNoiseRelates2023}. When SNR is low, the brain struggles to prioritize meaningful input, which negatively affects perceptual accuracy, decision-making, working memory \cite{anticevicNegativeNonemotionalInterference2011, starcSchizophreniaAssociatedPattern2017}, age-related working memory decline \cite{waschkeStatesTraitsNeural2017, voytekAgeRelatedChanges12015}, and motor function variability \cite{carmentNeuralNoiseCortical2020}. 

Multiple EEG studies report that individuals with schizophrenia exhibit decreased SNR \cite{abbasiIncreasedNoiseRelates2023, wolffItsTimingReduced2022}, especially in the prefrontal cortex \cite{wintererSchizophreniaReducedSignaltonoise2000, wintererGenesDopamineCortical2004}; with some research indicating that SNR is primarily due to increased neural noise rather than a deficit in stimulus-driven signal \cite{abbasiIncreasedNoiseRelates2023}. Because of the altered SNR, several studies suggest that neural noise metrics may serve as more reliable biomarkers for schizophrenia than traditional behavioral or oscillatory measures \cite{petersonAperiodicNeuralActivity2023}. Similarly, SNR metrics can reliably differentiate patients from healthy controls \cite{abbasiIncreasedNoiseRelates2023, wolffItsTimingReduced2022}. Our results show that neural noise may in fact be a key part of schizophrenia pathology.

\subsection{Contributions of this study}

In this study, we alter a deep reinforcement learning algorithm to simulate the excessive synaptic pruning and excitation-inhibition (E/I) imbalance that is associated with schizophrenia. Interestingly, we find that an E/I imbalance is insufficient to significantly alter the behavior of the learning algorithm, but that the \textit{combination} of E/I imbalance and additive noise induces a range of schizophrenia-like behaviors. This combination also reduces the sensitivity of the artificial neural network to reward-prediction-error (i.e. phasic dopamine). \textbf{Thus, this computational model suggests E/I imbalance + noise as conditions key for schizophrenia, and under which neurodevelopmental and dopamine hypotheses of schizophrenia can be reconciled.}

\section{Methods}

\subsection{Simulated environment}
We utilized a minigrid-style \cite{chevalier-boisvertMinigridMiniworldModular2023} goal-seeking task illustrated in figure 1, which places an agent in a small gridworld which it must learn to navigate. Three types of objects are present in the space: several “optional” goals (small reward), one “required” goal (large reward and ends the current episode), and a “hazard” (negative reward). Optional goals and hazards appear randomly throughout the room in each episode. At each step the agent chooses one of three actions: turn left, turn right, or move forward. The agent’s visual field only covers part of the room, so it must learn how to select appropriate actions to maximize reward, given the incomplete visual information. Both “healthy” and “schizophrenic” agents (with simulated excitation/inhibition imbalance) were placed in this environment and allowed to learn behavioral strategies.

While this simulation presents a goal-seeking problem involving spatial navigation, it is also a simple metaphor for the sequential decision-making of daily life: The optional goals represent opportunities like play, exploring interests, or socialization, while the required goal represents basic survival strategies such as finding food or employment, which must be attended to every day.

\subsection{Deep reinforcement learning model}

Each agent in our experiments is driven by a Deep Q Learning Algorithm \cite{mnihHumanlevelControlDeep2015} and features a feedforward perceptron-style artificial neural network using sigmoid activation functions such that neuron outputs range from 0 (i.e. fully quiet) to 1 (i.e. fully excited). The network has an input layer of 100 neurons that accept visual input, a hidden layer of 25 neurons, and an output layer of 3 neurons representing the 3 actions available to the agent at any given time: turn left, turn right, and move forward. Outputs estimate the values of each possible action from the current (visual) state. “Value” here is defined using the formulation common in temporal-difference learning \cite{suttonReinforcementLearningIntroduction2018}:

\begin{equation}
V \left ( s, a \right ) =  \mathbb{E} \left [  r + \gamma \cdot \max_{b \in A} \left [  V \left ( s', b \right ) \right ]  \right ]
\end{equation}

Where \textit{V(s,a)} is the value of executing action a while in state \textit{s, r} is the immediate reward received for that action (rewards can be positive or negative), \textit{s’} is the new state perceived after executing action \textit{a}, and $\gamma$ is a discount factor (between 0 and 1) that discounts the value of future rewards relative to immediate ones. After each experience in the environment, a reward-prediction-error (denoted $\delta$) is computed, capturing the difference between actual experienced value and the network’s estimate:

\begin{equation}
    RPE = \delta = r + \gamma \cdot \max_{b \in A} \left [  V \left ( s', b \right ) \right ] - V \left(s, a \right )
\end{equation}

It is thought that dopamine neurons in the midbrain signal RPE in the brain \cite{montagueFrameworkMesencephalicDopamine1996, schultzNeuralSubstratePrediction1997} and influence plasticity in the striatum \cite{schultzNeuralSubstratePrediction1997} (and possibly hippocampus \cite{mehrotraAccountingMultiscaleProcessing2023} and prefrontal cortex \cite{dawUncertaintybasedCompetitionPrefrontal2005}). In Deep RL, connection weights $w_i$ in the network are updated to reduce the RPE for next time (this simulates the neuromodulatory effect of dopamine):

\begin{equation}
    w_{i_{new}} = w_{i_{old}} - \alpha \frac{\partial \delta^2}{\partial w_{i_{old}}}
\end{equation}

where $\alpha$ is a learning rate parameter.

\subsection{Altering the network to simulate excitation/inhibition imbalance}

Our artificial neural network is randomly initialized with both positive (excitatory) and negative (inhibitory) connections. We simulate over-pruning of excitatory connections by selecting a random subset of positive connections and setting their weights to zero. The commonly-used backpropagation-based training process for neural networks can tune weights upward or downward, and so could simply replace the lost excitatory connections. To prevent this and maintain the simulated excitation/inhibition imbalance, we suppress the formation of strong positive connections using a selective weight decay effect. That is, we change the weight updates from the form in equation 3 to:

\begin{equation}
    w_{i_{new}} = w_{i_{old}} - \alpha \frac{\partial \delta^2}{\partial w_{i_{old}}} - \alpha\lambda \cdot \textup{RELU}(w_{i_{old}})
\end{equation}

where $\lambda$ is a parameter that effectively controls the magnitude of the resulting excitation/inhibition imbalance. The net effect of these alterations is a reduction in the overall amount of excitation relative to the amount of inhibition in the network. This is an \textit{ad-hoc} approach to creating an E/I imbalance in our artificial neural network, and does not necessarily reflect the particular pathology of E/I imbalance in biology.

For comparison, we also created an agent with the opposite excitation/inhibition imbalance. That is, we overpruned and penalized negative (inhibitory) connections to create an excess of excitation in the network. Comparison between the two types of imbalance help build validity for our computational model (see Results).

\subsection{Probing agents’ perceptions of their environment}
To partially understand how each agent perceives or processes its visual input, we create a second artificial neural network in parallel to the network used in decision making. This second network accepts the first network’s hidden layer neuron activations as inputs, and is trained to reconstruct the original visual input given these hidden layer signals.

The agent’s decision-making network processes visual input with the ultimate goal of executing rewarding behavior. The reconstructions of the visual input reveal something about this processing. For example, inaccurate reconstructions may accompany general impairments in goal-seeking behavior, or we may see reconstructions that primarily support goal seeking in some agents but reconstructions that primarily support hazard-avoidance in other agents.

\section{Results}

\subsection{The effects of simulated excitation/inhibition imbalance, and of additive noise}

After we apply the simulated excitation/inhibition imbalance, we allow the agent to learn in the environment until its behavior stabilizes, then measure all connection weights and neuron biases within the network. As expected, our intervention creates an excitation/inhibition imbalance that persists throughout learning. Interestingly, we found that this imbalance alone has very little effect on the agent’s performance, relative to an unaltered (“healthy”) agent. This is because the network can compensate for the general lack of excitation by increasing neuron biases (Fig 1). This increases general excitability within the network, allowing it to function even with lower overall levels of excitatory signalling. Thus the increase in neuron biases is a homeostatic mechanism employed by the artificial neural network. It should be noted that a biological network may try to achieve homeostasis through a different mechanism, but the \textit{principle} that the network attempts to compensate for the loss of excitation is likely general.

Importantly, we find that the increased excitability caused by the increased neuron biases makes the network less noise-tolerant. We add normally distributed random noise ($\sigma$ = 0.3) which has a roughly flat power spectrum, to the network’s inputs to simulate the increased noise observed in schizophrenia. This affected the excitation/inhibition-imbalanced network more than the unaltered network (Fig 1). Specifically, noise has the greatest effect on (all) networks’ ability to obtain optional rewards, but this effect is more dramatic for the network with excitation/inhibition imbalance. The more basic behavior of reaching the goal in each episode is preserved, suggesting a reversion from reward-rich behavior to simple survival strategy that is more pronounced in the network with excitation/inhibition imbalance.

\begin{figure}
  \centering
  \fbox{\includegraphics[width=\textwidth]{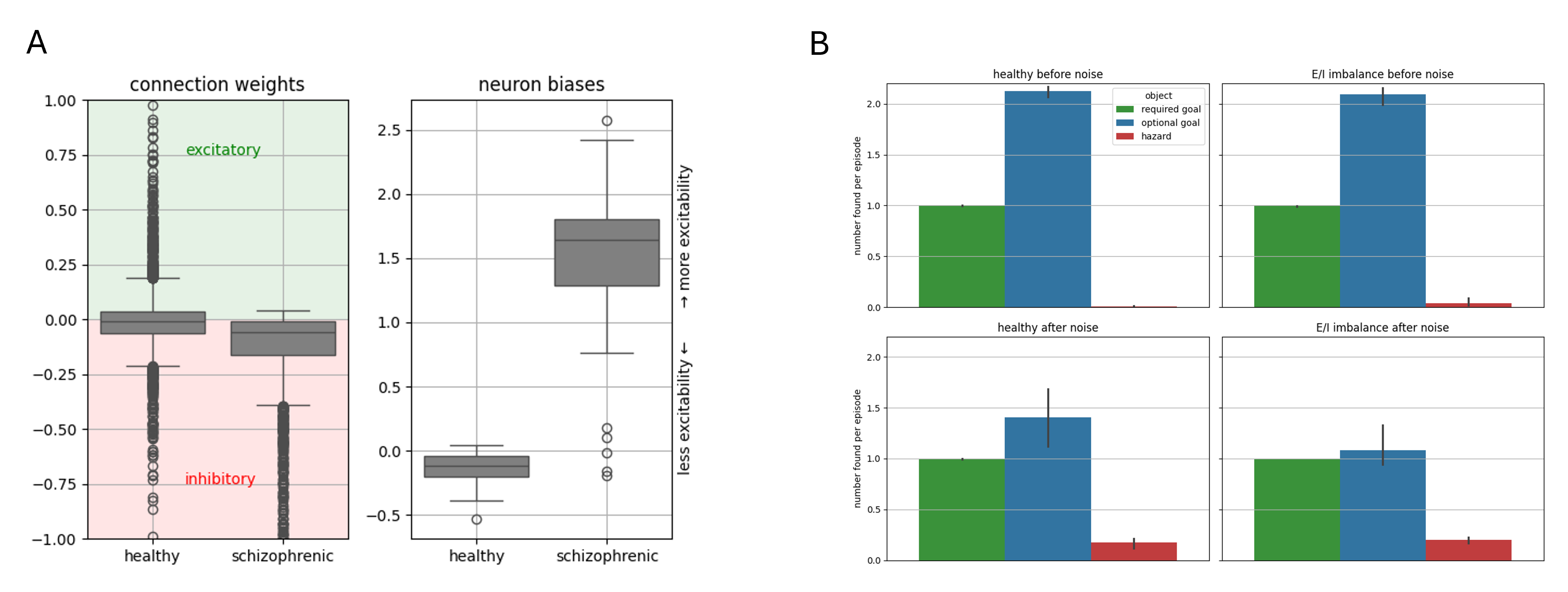} }
  \caption{A) Simulating excitation/inhibition imbalance creates an artificial neural network with mostly inhibitory connections. Interestingly, this has little effect on the agent’s behavior and performance, because the artificial network automatically compensates by increasing the bias of most neurons. This makes the neurons more “excitable”. B) Increased bias makes the network with excitation/inhibition imbalance less noise-tolerant. Here, adding gaussian noise to the network inputs causes all agents to obtain fewer optional rewards per episode, but the effect is greater for the agent with the imbalanced network.
}
\end{figure}

\subsection{Excitation/inhibition imbalance plus noise creates symptoms of schizophrenia}

The combination of excitation/inhibition imbalance and additive noise creates behavioral effects analogous to symptoms of schizophrenia, including anhedonia, avoidance, increased temporal discounting, and catatonia.

Figure 2 illustrates an anhedonia-like effect in which the schizophrenic agent appears ambivalent towards optional rewards. In a scenario analogous to a sucrose preference test \cite{bessaStressinducedAnhedoniaAssociated2013, zhuangTreadmillExerciseReverses2019}, we place agents in a position where the reward-optimal strategy would detour to collect an optional reward en route to the required goal - requiring 1 additional action but obtaining both rewards. The optimality of this detour is reflected in the “healthy” agent’s perceived values, but not the “schizophrenic” agent’s. In addition to this anhedonia-like effect, we see the avoidance-like effect illustrated in figure 2 where the schizophrenic agent seems to inappropriately generalize the negative value of a hazard to states where the hazard is not imminent.

\begin{figure}
  \centering
  \fbox{\includegraphics[width=\textwidth]{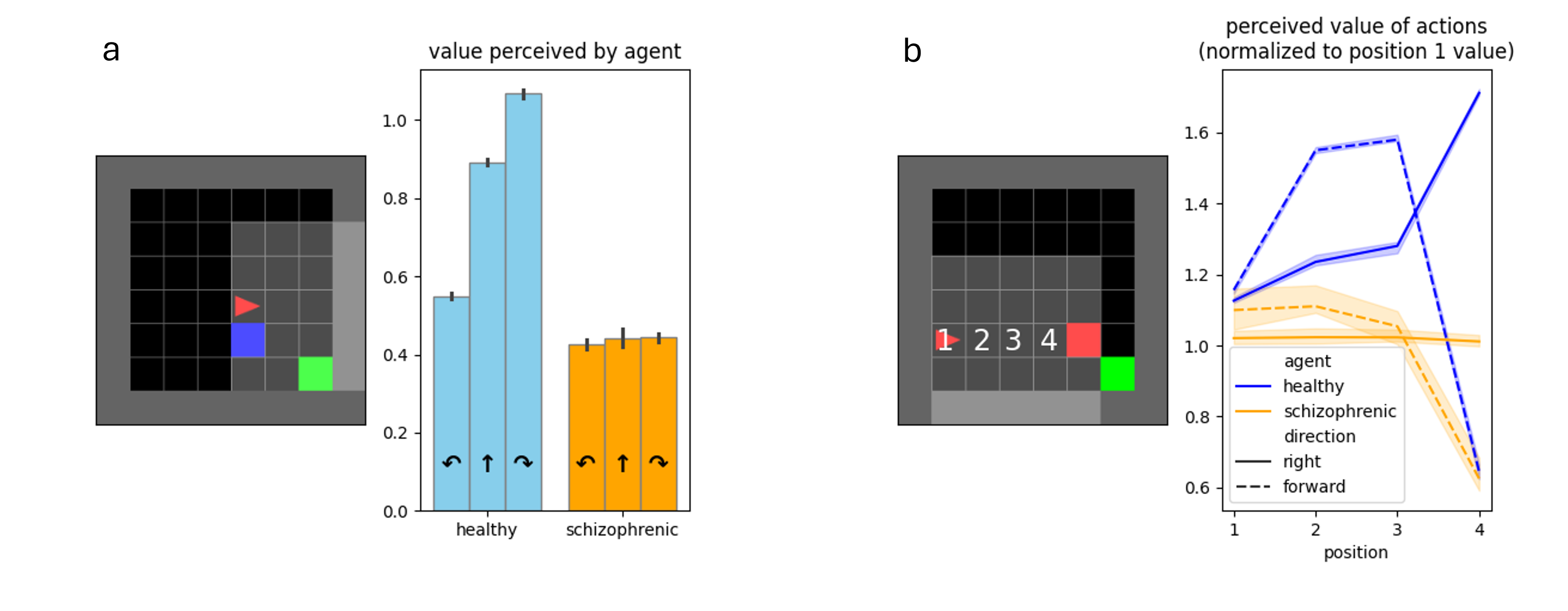} }
  \caption{a) In this contrived situation, reward-optimal behavior is to turn right, spending one additional action but collecting the optional reward en route to the goal. This is reflected in the “healthy” agent’s perceived action values, but the “schizophrenic” agent is ambivalent toward turning right or moving forward - an anhedonia-like effect. b) A contrived situation in which the agent is slowly moved toward a hazard. The “healthy” agent is appropriately willing to move forward through states 1-3, as this moves it closer to the goal. The “schizophrenic” agent demonstrates avoidance by devaluing forward moves much earlier.
}
\end{figure}

We can also infer the agent’s effective temporal discounting rate from changes in its action values as it approaches a goal or hazard. In our experiments, all agents use an explicit discount factor setting of $\gamma=0.9$, and the discount factor inferred from the “healthy” agent's behavior is roughly 0.9 as expected. Surprisingly, the inferred discount factor for the “schizophrenic” agent is roughly 0.75. That is, the excitation/inhibition imbalance (with additive noise) induces an additional discounting effect similar to that observed in schizophrenia \cite{amlungDelayDiscountingTransdiagnostic2019}. 

Finally, fig 3 illustrates a repetitive-movement effect analogous to stereotypy (a type of catatonia associated with schizophrenia). As the magnitude of the additive noise increases, the “schizophrenic” agent is increasingly likely to become “stuck” in particular states - executing the same movements repeatedly for some time before finally breaking free.

\begin{figure}
  \centering
  \fbox{\includegraphics[width=\textwidth]{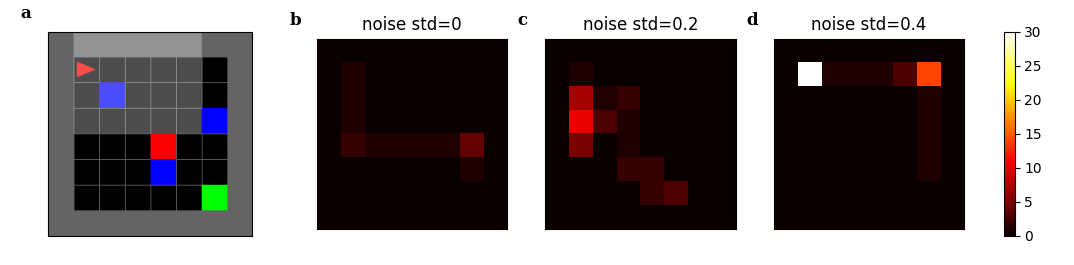} }
  \caption{Excitation/inhibition imbalance plus noise creates repetitive movements or catatonia. A) simulated environment. B-D) example heatmaps of how long (number of simulation steps) the agent spends in each state of the environment shown in A.
}
\end{figure}

\subsection{Inaccurate perception and reconstruction of the agent’s surroundings}
Immediately after adding noise to each agent’s neural network, we find that we can reconstruct the “healthy” agents’ visual input more accurately than the “schizophrenic” agents’. This suggests that the excitation/inhibition imbalance causes the network’s internal representations to be more easily disrupted in the presence of noise. The result is an effect analogous to hallucinations - with the schizophrenic agent’s reconstructions being a less accurate picture of reality, and more likely to contain false objects. Importantly, the “healthy” agents’ reconstructions are not perfectly accurate either: Figure 4 shows an example in which the healthy agent’s reconstruction is flawed, but still supports rewarding behavior (drawing the agent toward the goal in the top-left). To paraphrase Beau Lotto; the brain does not see the world as it is, but rather the world it is useful to see.

\begin{figure}
  \centering
  \fbox{\includegraphics[width=\textwidth]{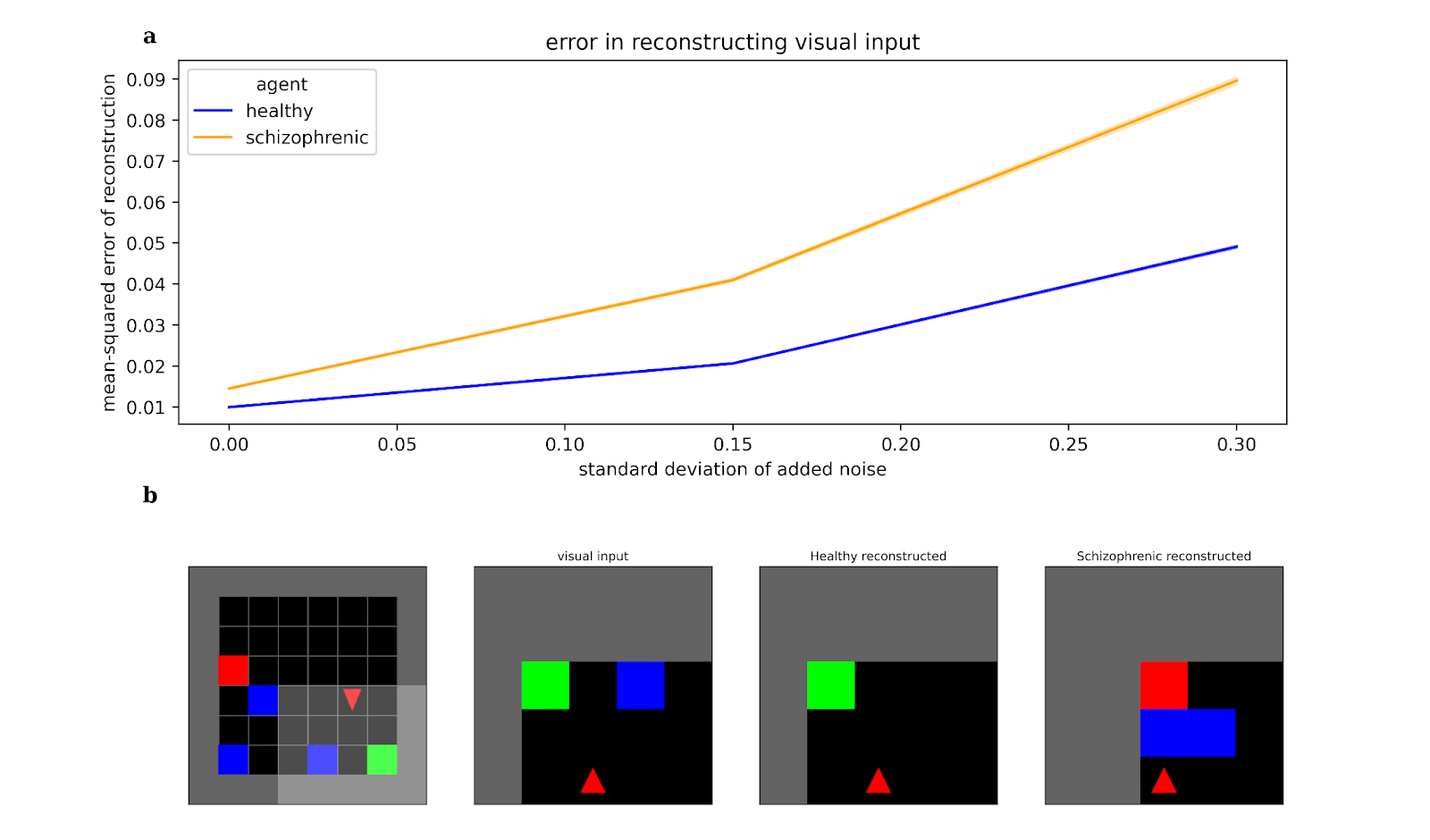} }
  \caption{a) when hidden-layer neuron activations are used to reconstruct the agents’ visual input, the “healthy” agent’s reconstructions are more accurate than the “schizophrenic” agent’s. This effect grows as the magnitude of additive noise increases. b) example reconstructions: neither agent reconstructs its surroundings perfectly, but the schizophrenic agent’s reconstruction is more hallucination-like (including an imagined hazard), and less likely to support rewarding behavior.
}
\end{figure}

\subsection{Altered response to reward-prediction error}
Phasic activity of dopamine is thought to signal reward prediction error, and has a neuromodulatory effect. Fig 9 shows the magnitude of the network’s response to reward prediction error (i.e. dopamine) in terms of mean weight change in the network per unit of reward prediction error. A given reward prediction error generally induces less change in the agent with simulated schizophrenia than the “healthy” agent. This suggests reduced plasticity and reduced sensitivity to reward prediction error in the schizophrenic network.

\begin{figure}
  \centering
  \fbox{\includegraphics[width=\textwidth]{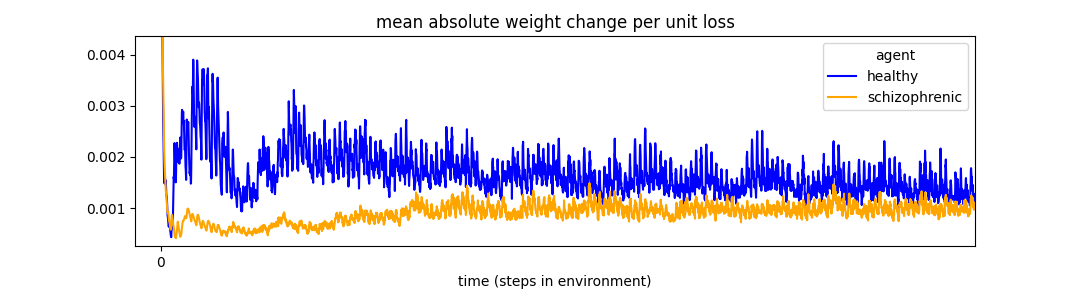} }
  \caption{Weight changes induced in the networks per unit loss (i.e per unit of reward-prediction error, which is thought to be signalled by dopamine in biological networks). The healthy network exhibits greater and sustained plasticity throughout learning. The “schizophrenic” network’s response to RPE is muted.
}
\end{figure}

\subsection{The opposite excitation/inhibition imbalance does not impair the agent}
To validate the specificity of our proposed model as a model of schizophrenia, we must consider whether the observed behavioral effects are specific to schizophrenia (i.e. is the agent impaired \textit{generally} or in specifically schizophrenia-like ways?), and whether the impairments are caused by our specific network alterations (i.e. would altering the network another way produce the same impairments?).

Our “schizophrenic” agent does assign value to reaching the goal, and does in fact reach the goal in each episode. This learning and execution of a basic survival strategy shows that the agent is not \textit{generally} impaired. In addition, when we reverse the excitation/inhibition imbalance to create a network with an \textit{excess} of excitation, we do not see any of the above impairments. In fact, the agent with reversed excitation/inhibition imbalance actually showed slightly enhanced performance relative to the “healthy” agent in terms of reward obtained in the presence of noise. Thus the schizophrenia-like behaviors we observed seem to depend specifically on reduced network excitation, plus noise.

\section{Discussion}
By creating a deep reinforcement learning agent with relatively more inhibition than excitation in its neural network and adding noise, we have created a computational model of schizophrenia with a degree of face validity. The deep reinforcement learning agent demonstrates behaviors such as anhedonia, increased temporal discounting, repetitive movement, and an inaccurate perception of its environment analogous to hallucinations - all features of schizophrenia. It still manages to execute a basic goal-seeking survival strategy, indicating that it is not generally impaired, and an opposite excitation/inhibition (E/I) imbalance does not produce the same schizophrenia-like behaviors, indicating that reduced excitation within the network is the specific deficiency necessary.

This model suggests a way to reconcile the dopamine hypothesis of schizophrenia with the neurodevelopmental hypothesis. In our model, the E/I imbalance seems to cause a reduction in the network’s sensitivity to reward-prediction-error (which in biological networks is signalled by dopamine). That is, the dopamine signal in our “schizophrenic” agent does not affect the same plastic change in the neural network that we see in our “healthy” agent. It is possible that dopamine fluctuations have reduced efficacy in a network which must - in addition to learning rewarding behavior - also expend energy to compensate for E/I imbalance. With the network’s sensitivity to dopamine reduced, the dopamine system may be thrown into dysregulation. Under this view, dopamine system dysregulation is not a cause of schizophrenia; rather, altered connectivity causes both schizophrenia and dopamine system dysregulation. Thus this model accounts for both E/I imbalance and dopamine system dysfunction, and lends computational support to the integrated hypothesis of schizophrenia.

However, we found that an excitation/inhibition imbalance is not sufficient to create behavioral changes - the network simply compensates for the reduced excitation. But in compensating for the imbalance, our network becomes more susceptible to noise. The addition of noise then produces the expected behavioral changes. Thus we propose the integrated hypothesis is not complete: the idea of noise as an essential ingredient should be added. We further suggest that treatments for schizophrenia should target either the E/I imbalance or noise processes in the brain, since the combination of E/I imbalance and noise seems to be prerequisite for schizophrenic symptoms in our model.

Our simulations model neural noise by adding normally-distributed random noise to the network’s input. Due to the feed-forward network nature of our artificial neural network, the noise then propagates through each layer and affects processing throughout the network. But this leaves some important questions open. What role does noise play, exactly, in the schizophrenic brain? Where does the noise come from, and does it exist at the level of individual-neuron function, network communication, or both? Can the noise be reduced through pharmacological, behavioral, or other means? Our artificial neural networks are a high-level abstraction of biological neural networks, with limited ability to comment on lower-level biological mechanisms. Just as our computational work addresses a limitation of biological studies - namely, the difficulty of controlling and studying E/I imbalance directly - the limitations of \textit{computational} work mean that we must now return to biology to seek more clarity on the sources and role of noise in schizophrenia.

Finally, it is interesting that the presence of noise turned out to be a necessary ingredient in a computational model. This is instructive for the field of computational modelling generally, where we usually simulate pristine signals and clean, precise neural processing. Including noise processes in our models will make them more realistic. As we have found here, the noise may actually be an important part of the reality of biological information processing - both its function and its dysfunctions.

\begin{ack}
The authors would like to acknowledge funding from Alberta Innovates, the Natural Sciences and Engineering Research Council of Canada (NSERC), and Mount Royal University.
\end{ack}

{
\small
\bibliographystyle{unsrt}
\bibliography{library, MDD-AI-MRU}

}

\end{document}